\documentclass[conference]{IEEEtran}
\IEEEoverridecommandlockouts
\usepackage{cite}
\usepackage{amsmath,amssymb,amsfonts}
\usepackage{algorithmic}
\usepackage{graphicx}
\usepackage{textcomp}
\usepackage{xcolor}
\def\BibTeX{{\rm B\kern-.05em{\sc i\kern-.025em b}\kern-.08em
    T\kern-.1667em\lower.7ex\hbox{E}\kern-.125emX}}
\begin{document}

\title{On the Performance of ISAC over the D-Band in a Phase-Noise Aware OFDM Systems 
}

\author{\IEEEmembership{}
        \IEEEmembership{}        
\author{Didem Aydoğan, Mohaned Chraiti,  Korkut Kaan Tokgöz\\
Faculty of Engineering and Natural Sciences, Electronics Engineering Department\\
Sabancı University, İstanbul, Türkiye \\
\small{ didem.aydogan@sabanciuniv.edu, mohaned.chraiti@sabanciuniv.edu, korkut.tokgoz@sabanciuniv.edu }}}

\maketitle
\begin{abstract}
Phase noise (PN) is a critical impairment at D-band frequencies (110–170 GHz), which are widely investigated as promising candidates for beyond-5G/6G ISAC systems. This paper evaluates OFDM-based ISAC sensing performance under realistic oscillator impairments using a hardware-tuned 3GPP PN model at 130 GHz and FFT-based radar processing. With a numerology of 480 kHz ($\mu$ = 5), results show that PN introduces range RMSE floors of 0.04–0.05 m and velocity RMSE floors of 0.12–0.18 m/s. Doppler sidelobe metrics also saturate, with PSLR around –6 dB and ISLR around –4 dB.

These findings confirm that range accuracy remains bandwidth-limited, while velocity estimation and sidelobe suppression are strongly PN-sensitive. The study highlights the importance of PN-aware waveform and numerology design for sub-THz ISAC and provides insights for future multi-band transceivers. Communication metrics and PN mitigation strategies such as PTRS and CPE tracking are left for future work.

\end{abstract}




\section{Introduction}
Integrated Sensing and Communication (ISAC) is regarded as a core technology in sixth-generation (6G) wireless networks, aiming to seamlessly combine high-resolution sensing with high-speed data transmission to enable applications such as autonomous driving, extended reality (XR), and industrial automation \cite{wild2021ieee, saad2020overview}. Meanwhile, 6G systems are anticipated to operate over an extended frequency range, including less explored bands, particularly within the millimeter-wave (mmWave) and sub-terahertz (sub-THz) spectrum spanning 24–300\,GHz \cite{towards_6G_AVVR, above100_isac}. Although current 3GPP standardization efforts (FR2/FR3) do not yet include the D-band, it has been widely investigated as a promising candidate for beyond-5G/6G due to its potential for very large bandwidths and fine spatial resolution.  

\par While mmWave bands (24-71\,GHz) have demonstrated the early feasibility of ISAC, moving into the sub-THz spectrum introduces new challenges. Operation at such high frequencies exacerbates propagation losses, reduces radar cross-section (RCS), and increases susceptibility to hardware impairments, most notably oscillator phase noise (PN). These impairments can compromise both sensing accuracy and communication reliability, especially in monostatic ISAC systems where the same transceiver and waveform are shared across functions \cite{fundamental_RCS}.  

A central technical question in monostatic ISAC is the choice of waveform. Orthogonal Frequency Division Multiplexing (OFDM) is a widely studied and practical option thanks to its radar–communication compatibility, high spectral efficiency, and seamless integration with MIMO processing \cite{mmwaveOFDM_radar, mimo_ofdm_ISAC}. Prior studies have shown that in the absence of hardware impairments, OFDM can achieve accurate range–Doppler estimation while supporting unified waveform optimization for joint sensing and communication \cite{mimo_ofdm_ISAC}. Nonetheless, it should be noted that alternative schemes such as OTFS and delay–Doppler domain techniques are also under active investigation, particularly for high-mobility environments.  

\par As communication shifts to higher frequencies, circuit limitations and hardware impairments—particularly PN—become more pronounced, degrading OFDM performance. At sub-THz, PN introduces common phase error (CPE) and inter-carrier interference (ICI), disrupting subcarrier orthogonality \cite{understanding_effectsPN, PN_THEORICAL_model}. Standardized PN models (e.g., 3GPP, Hexa-X) provide useful baselines but often fail to capture realistic oscillator behavior in the D-band \cite{PN_modeling_real}. Practical mitigation methods such as PT-RS insertion and CPE tracking have proven effective at mmWave, but their robustness diminishes under higher PN variance and sparse pilot configurations \cite{8821684, 8690852}. In parallel, scalable OFDM numerology introduced in 3GPP Release~15 enables subcarrier spacing adaptation to diverse mobility and resolution needs \cite{subcarriier_modulation_specification, study_numerol_V2X}. In ISAC, numerology selection directly affects range and Doppler resolution, as well as resilience to PN. Wider spacing improves range resolution but reduces Doppler granularity, while narrower spacing enhances Doppler performance at the expense of increased PN sensitivity and computational cost. Effective velocity resolution further depends on integration time and SNR, meaning that higher frequencies alone do not guarantee superior Doppler accuracy.  

\par The sensing performance limits of OFDM-based ISAC under severe D-band PN remain underexplored. This paper addresses this gap by evaluating radar estimation under realistic oscillator impairments using a hardware-tuned 3GPP PN model at 130\,GHz \cite{Tunned_PN}. We assess 5G numerologies and identify $\mu = 5$ (480\,kHz) as a balanced choice, then quantify PN impact on range, velocity, and Doppler sidelobes via low-complexity FFT-based radar processing. Our study provides practical insights for PN-aware waveform design in sub-THz ISAC. Unlike prior works that rely on idealized PN models for communication \cite{Tunned_PN} or theoretical radar analysis \cite{keskin2022ofdm}, this paper combines a hardware-calibrated D-band PN model with numerology-aware FFT radar processing, enabling a realistic, implementation-ready evaluation of ISAC performance.  

\par Our contributions are as follows:  

\begin{itemize}
    \item Simulate the OFDM-based ISAC radar at 70\,GHz and 130\,GHz using a hardware-tuned PN model.  
    \item Evaluate 5G numerologies and identify $\mu = 5$ as a balanced sensing configuration.  
    \item Quantify PN-induced degradation using RMSE, PSLR, and ISLR metrics.  
    \item Provide design insights for robust ISAC across multi-band and sub-THz systems.  
\end{itemize}  

\par Our findings emphasize the need for hardware-aware ISAC design at sub-THz frequencies and provide a foundation for evaluating sensing robustness across emerging 6G frequency bands. The rest of this paper is organized as follows. Sec.~\ref{sec:system_model} introduces the system and signal models, including PN and 2D FFT-based radar processing. Sec.~\ref{sec:num} discusses numerology trade-offs in range and Doppler resolution. Sec.~\ref{sec:res} presents the simulation setup and results. Sec.~\ref{sec:clc} concludes the paper and outlines future directions.

\section{System Model}\label{sec:system_model}

We consider a monostatic OFDM-ISAC system operating at a carrier frequency of $f_c = 130 \mathrm{GHz}$, with co-located transmit and receive antennas. The transmit waveform comprises $M$ OFDM symbols over $N$ subcarriers, each carrying a data symbol $x_{\ell,m}$ where $\{l,m\}\in[1,N]\times[1,M]$. This unified waveform allows simultaneous communication and radar sensing by exploiting OFDM’s frequency-domain orthogonality. 

Post cyclic prefix removal and an \(N\)-point DFT, the received signal at subcarrier \(\ell\) during symbol index \(m\) captures the effects of the target’s range delay \(\tau\), Doppler shift \(\nu\), oscillator impairments, and additive white Gaussian noise (AWGN).

\subsection{Transmitted OFDM Signal}
The transmitter generates a block of \( M \) consecutive OFDM symbols, each consisting of \( N \) subcarriers with spacing \( \Delta f \). The frequency-domain data symbol on subcarrier \( \ell \) of the \( m \)th OFDM symbol is denoted by \( x_{\ell,m} \), and is drawn from a 16-QAM constellation. An \( N \)-point inverse fast Fourier transform (IFFT) is applied to the vector \( \mathbf{x}_m = [x_{0,m},  x_{1,m},  \dots,  x_{N-1,m}]^T \), yielding the corresponding time-domain waveform.

\[
 s_m[n]=\frac{1}{\sqrt{N}}\sum_{\ell=0}^{N-1}x_{\ell,m}e^{j2\pi\ell n/N},\quad n=0,\dots,N-1.
\]
A cyclic prefix of length \( N_{\mathrm{CP}} = N/4 \) is appended to each OFDM symbol to construct the transmitted waveform \( s^{\mathrm{tx}}_m[n] \), thereby preserving subcarrier orthogonality and mitigating inter-symbol interference caused by multipath propagation.

\subsection{Signal Model with Phase Noise}\label{sec:pn_model}
At sub-THz frequencies, oscillator imperfections, referred to as PN, constitute a major impairment. PN arises from local oscillator (LO) instabilities during frequency synthesis and is aggravated by the high multiplication factors required at sub-THz bands. In OFDM-based sensing, PN disrupts subcarrier orthogonality, leading to both CPE and ICI, which in turn degrade Doppler estimation and range accuracy. While the impact of PN on communication performance has been extensively studied, this work focuses on its influence on sensing accuracy in the sub-THz regime. Furthermore, CPE correction typically depends on the density of pilot carriers, which may introduce additional errors in target parameter estimation.

Post cyclic prefix removal and $N$-point DFT processing at the receiver, the baseband observation on subcarrier $\ell$ during OFDM symbol $m$ can be expressed as:
\begin{equation}
y_{\ell,m} = x_{\ell,m} \cdot e^{-j2\pi \ell \Delta f \tau} \cdot e^{j2\pi f_c m T_s \nu} \cdot w_{\ell,m} + z_{\ell,m},
\label{eq:rx}
\end{equation}
where $\tau$ and $\nu$ are the sensed target’s range delay and Doppler shift, respectively, and $T_s = 1/\Delta f$ is the OFDM symbol duration (excluding cyclic prefix). The received signal is affected by the target’s RCS, which is implicitly included in the signal amplitude. Typical RCS values at D-band are 10–20 dBsm for vehicles and 0–5 dBsm for pedestrians \cite{fundamental_RCS}. Particularly Indoor scenarios requires analysis for lower RCS (typical -20 dBsm for a drone), corresponding to small object detection. 

The term $w_{\ell,m} = \exp\big(j[\phi(t_m - \tau) - \phi(t_m)]\big)$ captures the differential PN between transmission and reception. The additive term $z_{\ell,m} \sim \mathcal{CN}(0,\sigma^2)$ models AWGN. The oscillator PN process $\phi(t)$ is modeled as a zero-mean Gaussian random process, characterized by its one-sided power spectral density (PSD). In this work, two PSD profiles are considered:
\begin{itemize}
    \item 3GPP Model at 70 GHz: A standard reference model for mmWave systems.
    \item Hardware-tuned model at 130 GHz: Fitted to measured oscillator data to reflect realistic D-band transceivers \cite{Tunned_PN}.
\end{itemize}

From each PSD, discrete-time PN sequences $\phi[n]$ are synthesized at the system sampling rate $f_s = N \Delta f$, and used to compute:
\[
w_{\ell,m} = \exp\big(j[\phi[n(m - \tau)] - \phi[nm]]\big),
\]
which multiplicatively impairs each subcarrier.

Tab. \ref{tab:pn_compare} compares parameters of the standard 3GPP PN model at 70 GHz and a hardware-tuned version calibrated to 130 GHz oscillator measurements \cite{Tunned_PN}. Both models follow a similar structural format, but the tuned model exhibits higher reference noise levels and elevated 1/$f$ and 1/$f^3$ corner frequencies. These shifts reflect the more severe phase instability encountered at sub-THz frequencies due to oscillator limitations and frequency multiplication effects.

\begin{table}[h]
\centering
\caption{Phase Noise Model Parameters}
\label{tab:pn_compare}
\begin{tabular}{|l|c|c|}
\hline
\textbf{Parameter} & \textbf{Tuned (130 GHz)} & \textbf{3GPP (70 GHz)} \\
\hline
$PLL_{ref}$ Noise [dBc/Hz] & –70 & –39.5 \\
White Noise [dBc/Hz]         & –150 & –111 \\
1/$f$ Corner $f_{c,1}$ [Hz]   & $1.1 \times 10^4$ & $3.1 \times 10^3$ \\
1/$f^3$ Corner $f_{c,2}$ [Hz] & $1.1 \times 10^7$ & $3.96 \times 10^5$, $7.54 \times 10^8$ \\
\hline
\end{tabular}
\end{table}

In practice, the hardware-tuned 130 GHz model exhibits significantly higher corner frequencies—$f_{c,1} = 1.1 \times 10^4$ Hz and $f_{c,2} = 1.1 \times 10^7$ Hz—compared to the 70 GHz 3GPP model, which uses lower values around $3.1 \times 10^3$ Hz and $3.96 \times 10^5$ Hz, respectively. These elevated corner frequencies indicate that the 130 GHz oscillator transitions into its 1/$f$ and 1/$f^3$ regions earlier and maintains high PN power over a wider offset range. As a result, more PN energy falls within the occupied OFDM bandwidth—especially in systems with narrow subcarrier spacing or long observation durations.

This spectral behavior increases the likelihood of ICI, degrades subcarrier orthogonality, and leads to elevated sidelobe levels in Doppler estimation. These effects are particularly pronounced when the OFDM symbol duration is long or the system relies on high-resolution Doppler processing, both of which make the receiver more sensitive to phase drift. Hence, under the 130 GHz tuned model, PN-induced degradation becomes a key limiting factor in velocity estimation and sidelobe suppression.

\subsection{2D FFT Processing}\label{sec_2dfft}

To estimate target range and velocity, we apply a two-dimensional (2D) discrete Fourier transform (DFT) to the received OFDM signal after removing the cyclic prefix and compensating for the known transmit waveform. Let \(r[n,m]\) denote the baseband signal samples, where \(n = \{0, \dots, N-1\}\) indexes subcarriers (frequency/range domain) and \(m = \{0, \dots, M-1\}\) indexes OFDM symbols (time/Doppler domain).
The 2D FFT is defined as:
\[
R[k,\ell] = \frac{1}{\sqrt{NM}} \sum_{n=0}^{N-1} \sum_{m=0}^{M-1}
r[n,m]  e^{-j2\pi \left(\frac{kn}{N} + \frac{\ell m}{M}\right)},
\]
where \(k = 0, \dots, N-1\) and \(\ell = 0, \dots, M-1\) correspond to discrete range and Doppler frequency bins, respectively.

The resulting 2D spectrum \(R[k,\ell]\) forms the range–Doppler map, with energy peaks at bin \((k^*, \ell^*)\) corresponding to the most likely target parameters. These peaks are mapped to physical quantities using:
\begin{align}
\hat{\tau} &= \frac{k^*}{2 N \Delta f}, \\
\hat{\nu}  &= \frac{\ell^*}{2 fc M T_{\mathrm{int}}},
\end{align}
where \(T_{\mathrm{int}} = M T_s\) is the total coherent processing interval and $fc$ is the carrier frequency.

The FFT-based approach leverages the inherent structure of OFDM waveforms for joint sensing, enabling low-complexity estimation of target range and velocity. It forms the basis for computing the root mean square error (RMSE), peak-to-sidelobe ratio (PSLR), integrated sidelobe ratio (ISLR), and resolution metrics presented in the subsequent sections.

\section{System Configuration and Requirements}\label{sec:num}

The configuration of waveform parameters, such as bandwidth, subcarrier spacing, and modulation format, has a direct impact on the achievable range resolution and data rate in D-band ISAC systems. Tab.~\ref{tab:sensing} outlines representative design targets consistent with anticipated 6G applications operating in the 130–140 GHz range.

\begin{table}[h]
\centering
\caption{Representative Requirements for D-band ISAC Sensing and Communication}
\label{tab:sensing}
\begin{tabular}{|c|c|}\hline
\textbf{Parameter} & \textbf{Typical Range} \\ \hline
Carrier Frequency & 130–140 GHz \\ \hline
Bandwidth & 4–10 GHz \\ \hline
Subcarrier Spacing & 120 kHz – 1 GHz \\ \hline
Waveform Type & OFDM \\ \hline
Target Range Resolution & $ < 10 cm$ \\ \hline
Target Applications & Vehicular radar, XR/AR, Factory automation \\ \hline
\end{tabular}
\end{table}

As demonstrated in prior works \cite{wild2021ieee, sen2023sensors, keskin2022ofdm, wang2021dfts}, the configurations in the table reflect a balance between high-resolution sensing and robust communication under hardware constraints. For instance, a bandwidth of 10 GHz theoretically supports range resolutions as fine as 1.5 cm, while subcarrier spacing directly impacts both Doppler sensitivity and system resilience to oscillator PN. The trade-offs among these parameters drive the design space for practical sub-THz ISAC systems.

\subsection{OFDM Numerology Trade-Offs for ISAC}\label{sec:ofdmtradeoffs}

The performance of D-band ISAC systems is tightly coupled to the choice of OFDM waveform parameters, particularly subcarrier spacing $\Delta f$, the number of subcarriers $N$, and total bandwidth $B$. These parameters jointly determine the range and velocity resolution, system complexity, and sensitivity to oscillator impairments.

Tab.~\ref{tab:ofdm_tradeoff_fixedbw} summarizes key trade-offs for a fixed bandwidth of 1.5 GHz at $f_c = 130$ GHz, assuming $M=64$ OFDM symbols and an antenna aperture of $D = 0.1$ m. As the numerology index $\mu$ increases (doubling $\Delta f$), the number of active subcarriers $N = B/\Delta f$ decreases, shortening the OFDM symbol duration $T_s = 1/\Delta f$ and integration time $T_{\mathrm{int}} = M T_s$.

\begin{table}[h]
\centering
\caption{OFDM Parameter Trade-Offs with Fixed BW = 1.5 GHz (Assuming $M=64$, $f_c=130$ GHz, $D=0.1$ m)}
\label{tab:ofdm_tradeoff_fixedbw}
\begin{tabular}{|c|c|c|c|c|c|c|}
\hline
\textbf{Numerology} & $\Delta f$ & $N$ & $T_s$ & $\Delta R$ & $\Delta v$ & $\Delta\theta$ \\
$\mu$ & [kHz] &  & [$\mu$s] & [m] & [m/s] & [deg] \\
\hline
4 & 240   & 6250 & 4.17 & 0.10 & 0.044 & 0.13 \\
5 & 480   & 3125 & 2.08 & 0.10 & 0.177 & 0.13 \\
6 & 960   & 1562 & 1.04 & 0.10 & 0.710 & 0.13 \\
7 & 1920  & 781  & 0.52 & 0.10 & 2.84  & 0.13 \\
\hline
\end{tabular}
\end{table}

While the range resolution $\Delta R = c/(2B)$ remains constant due to fixed bandwidth, velocity resolution degrades as $T_{\mathrm{int}}$ shortens. Low numerologies (e.g., $\mu=4$) require high $N$ (6250), increasing FFT size and memory load which is impractical for real-time hardware. Higher numerologies ($\mu=6,7$) reduce computational burden but severely limit Doppler sensitivity.

\subsection{Numerology-Dependent Resolution Trade-Offs}

The subcarrier spacing in 5G NR and beyond follows the well-known numerology:
\begin{equation}
\Delta f = 15 \times 2^{\mu} \quad \text{[kHz]}, \quad \mu \in \{4,5,6,7\}.
\end{equation}
Fixing FFT size at $N=256$, each numerology yields an effective bandwidth $B_{\mu} = N \Delta f$. The corresponding range and velocity resolutions are given by:
\begin{equation}
\Delta R = \frac{c}{2 N \Delta f}, \qquad \Delta v = \frac{c \Delta f}{2 f_c M}.
\end{equation}

\begin{figure*}[t]
  \centering
  \includegraphics[width=1\textwidth]{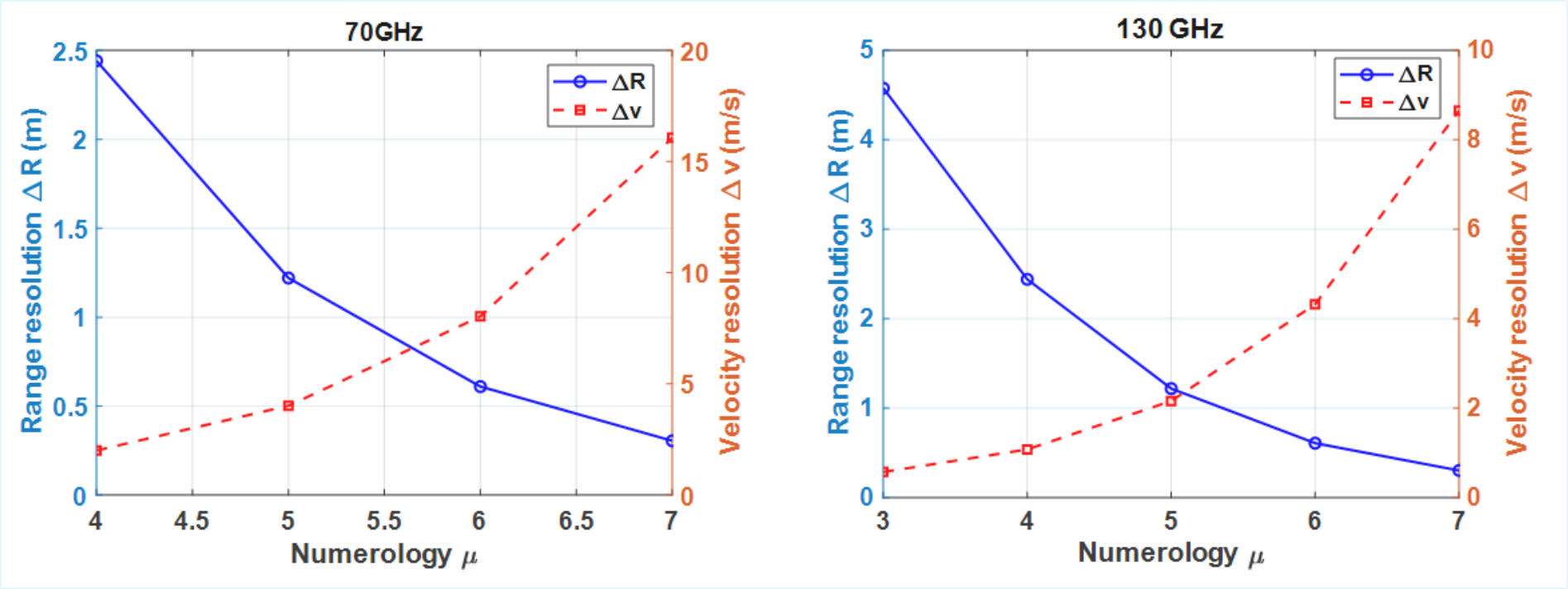}
  \caption{Range and velocity resolution at 70 GHz and 130 GHz as functions of numerology $\mu$ for $N = 256$.}
  \label{fig:res_vs_mu}
\end{figure*}

The trade-off is illustrated in Fig.~\ref{fig:res_vs_mu}. As $\mu$ increases, the range resolution improves due to wider subcarrier spacing, while velocity resolution degrades because of the shorter coherent integration time. For example, at 130 GHz, $\mu = 4$ yields $\Delta v \approx 0.044$ m/s and $\Delta R \approx 2.44$ m, whereas $\mu = 7$ provides fine range resolution ($\approx 0.305$ m) but coarse velocity resolution ($\approx 2.84$ m/s). Notably, the range and velocity resolution curves intersect around $\mu = 5$, indicating a balanced operating point where the system achieves moderate performance in both dimensions. This crossover makes $\mu = 5$ a practical numerology choice for D-band ISAC systems requiring both sufficient Doppler tracking and range discrimination under hardware constraints.

\subsection{Performance Metrics for ISAC Evaluation}

To comprehensively assess ISAC performance in the D-band, we consider the following performance metrics:

\begin{itemize}
  \item {Range Resolution} ($\Delta R$) and {Velocity Resolution} ($\Delta v$): primary radar performance indicators.
  \item {Peak and Integrated Sidelobe Ratios (PSLR, ISLR)}: reflect Doppler-domain interference and target separability.
  \item {Spectral Efficiency}: evaluates overall bandwidth utilization for data transmission.
\end{itemize}
These metrics reflect the trade-offs between sensing accuracy, signal robustness, and spectral efficiency in ISAC systems. They underscore the challenge of improving radar performance without compromising communication quality, particularly in sub-THz scenarios where phase noise and hardware constraints are significant. The selected numerology $\mu = 5$ achieves a balanced subcarrier spacing and symbol duration, preserving subcarrier orthogonality and ensuring reliable demodulation under PN. It also limits pilot overhead and processing latency, supporting efficient data transmission. Although this study focuses solely on sensing performance, the waveform configuration does not disrupt communication, validating its suitability for practical dual-function ISAC deployment.

\section{Simulation and Discussion}\label{sec:res}
We simulate a monostatic OFDM-based ISAC system operating at 130 GHz, aimed at jointly estimating the range and velocity of a radar target positioned 5 m away and moving at a speed of 1.5 m/s. The system employs a constant signal bandwidth of 1 GHz, yielding a theoretical range resolution of approximately 15 cm. Based on prior analysis (Fig.~\ref{fig:res_vs_mu}), a subcarrier spacing of $\Delta f = 480$ kHz (numerology index $\mu = 5$) is selected as a balanced operating point where range and velocity resolutions intersect. This numerology offers adequate Doppler sensitivity and range discrimination while maintaining a manageable FFT size for real-time hardware.

The simulation framework transmits known 16-QAM symbols over an OFDM time–frequency grid and employs FFT-based radar processing to estimate the target’s range and Doppler shift. The PN is modeled using a hardware-tuned 3GPP profile calibrated for 130 GHz oscillators, and performance is evaluated across various SNR levels. Key metrics include range and velocity RMSE, PSLR, and ISLR. Each SNR point is averaged over 200 Monte Carlo trials to ensure statistical reliability. An FFT size of 2048 is used to provide sufficient spectral resolution and oversampling, enabling precise peak localization in the range–Doppler map while remaining compatible with realistic radar processing hardware. The simulation parameters are summarized in Tab.~\ref{tab:sim_params}.

\begin{table}[h]
\centering
\caption{Simulation Parameters (Fixed $\Delta f = 480$ kHz)}
\label{tab:sim_params}
\begin{tabular}{|l|c|l|}
\hline
\textbf{Parameter} & \textbf{Symbol} & \textbf{Value} \\
\hline
Carrier Frequency & $f_c$ & 70 GHz, 130 GHz \\
\hline
Subcarrier Spacing & $\Delta f$ & 480 kHz \\
\hline
Numerology Index & $\mu$ & 5 \\
\hline
Number of Subcarriers & $N$ & 256 \\
\hline
FFT Size & $N_{\mathrm{FFT}}$ & 2048 \\
\hline
Cyclic Prefix Length & $N_{\mathrm{CP}}$ & 64 \\
\hline
OFDM Symbols per Frame & $M$ & 64 \\
\hline
Sampling Rate & $f_s$ & 122.88 MHz \\
\hline
Target RCS & - & $-20$ dBsm \\
\hline
\end{tabular}
\end{table}

\begin{figure*}[t]
\centering
\includegraphics[width=1\textwidth]{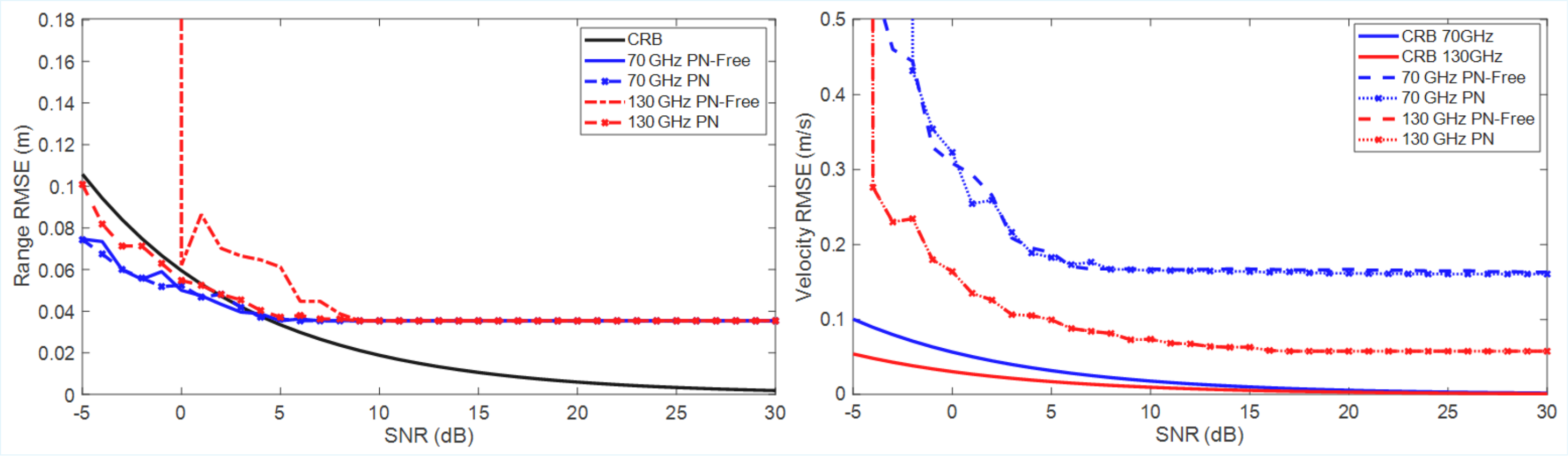}
\caption{Range and velocity estimation RMSE versus SNR for 70 GHz and 130 GHz carriers with fixed $\Delta f = 480$ kHz, shown both without PN (solid lines) and with hardware-tuned PN (dashed lines). Cramér–Rao bounds (CRBs) represent ideal performance limits.}
\label{fig:rmse_snr}
\end{figure*}

Fig.~\ref{fig:rmse_snr} shows the RMSE for range and velocity estimation versus SNR for both 70\,GHz and 130\,GHz carriers. In the range domain (left), RMSE decreases with SNR and follows the general trend of the CRB, but with a persistent gap at high SNR. The error floor appears at approximately 0.04-0.05\,m for both frequencies, caused by residual PN and finite FFT resolution. The PN-free and PN-impaired curves nearly overlap, confirming that range accuracy is dominated by bandwidth and only marginally affected by oscillator impairments.  

In contrast, velocity RMSE (right) is much more sensitive to both carrier frequency and PN. At high SNR, PN-free operation yields floors of $\approx 0.15$\,m/s at 70\,GHz and $\approx 0.08$\,m/s at 130\,GHz, reflecting the finer Doppler resolution at shorter wavelength. With PN included, the estimation floors rise to $\approx 0.18$\,m/s and $\approx 0.12$\,m/s, respectively. These results show that while 130\,GHz provides better intrinsic velocity resolution, its advantage is moderated by stronger PN sensitivity.  

\begin{figure}[t]
\centering
\includegraphics[width=0.48\textwidth]{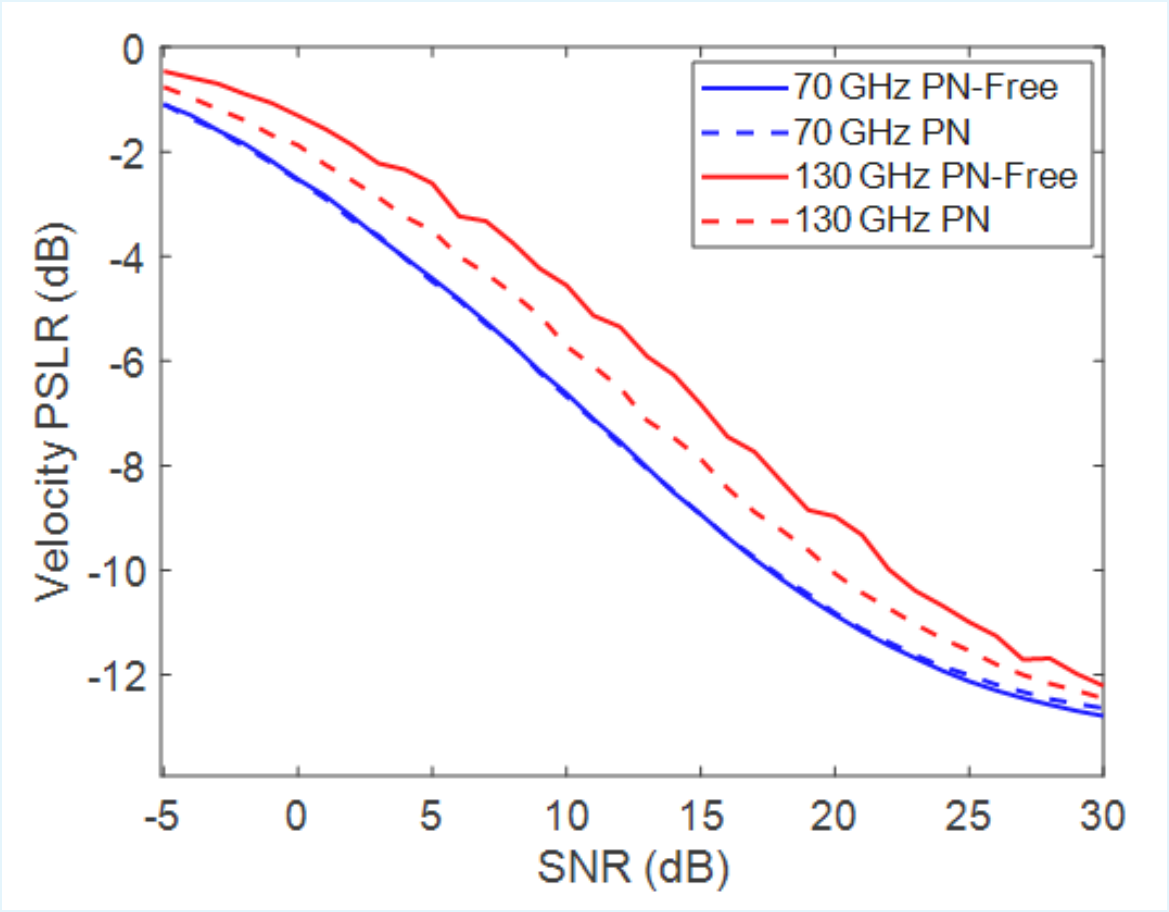}
\caption{Velocity-domain PSLR versus SNR at 130 GHz with and without PN. The PN limits sidelobe suppression, causing saturation at approximately –6 dB.}
\label{fig:pslr}
\end{figure}

\begin{figure}[t]
\centering
\includegraphics[width=0.48\textwidth]{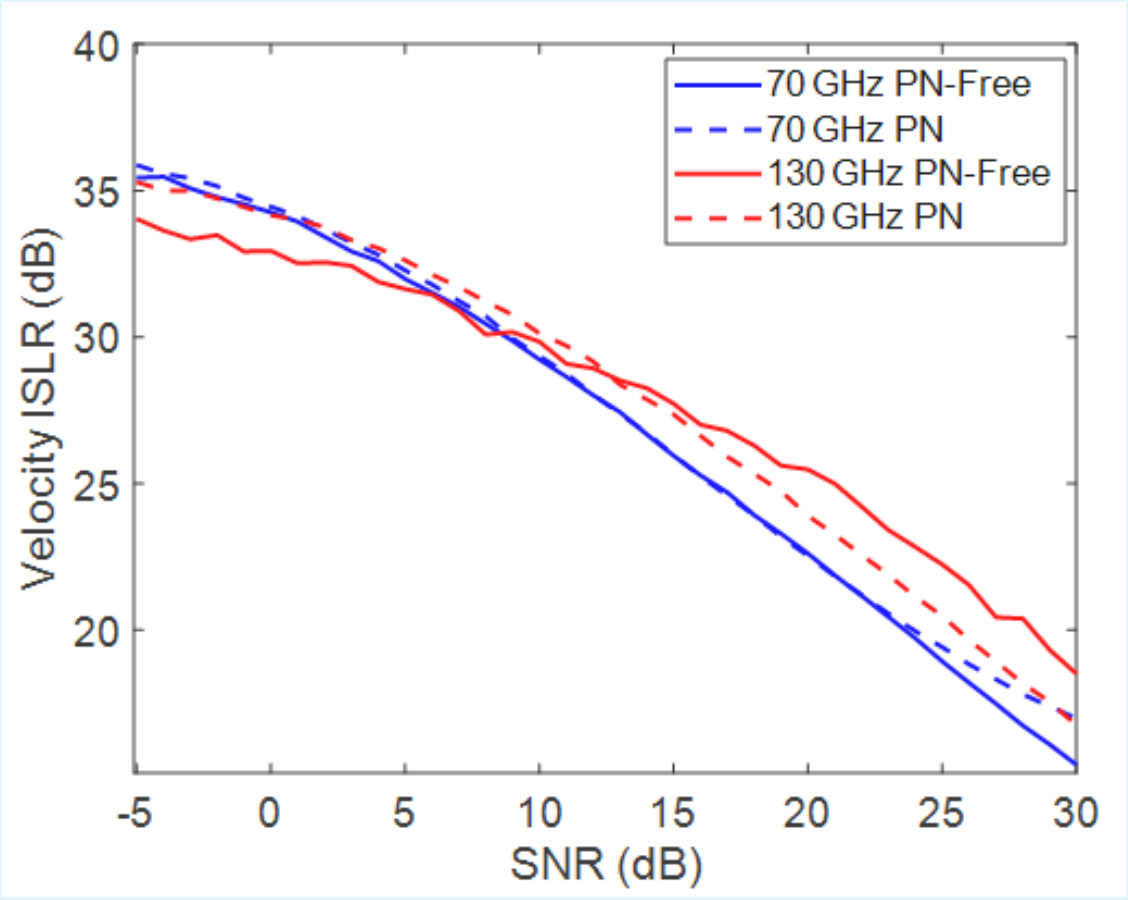}
\caption{Velocity-domain ISLR versus SNR at 130 GHz. The PN elevates background energy, with ISLR flattening above –4 dB.}
\label{fig:islr}
\end{figure}

To assess Doppler domain fidelity, Figs.~\ref{fig:pslr} and~\ref{fig:islr} report PSLR and ISLR performance. Without PN, sidelobe suppression improves with SNR, reaching below $-12$\,dB for PSLR and about $-7$ to $-8$\,dB for ISLR at 30\,dB SNR. Under PN, however, both metrics saturate, with PSLR limited to $\approx -6$\,dB and ISLR to $\approx -4$\,dB. This saturation restricts the system’s ability to separate closely spaced Doppler targets, even at high SNR.  

These trends are directly tied to PN model characteristics. The hardware-tuned model at 130\,GHz features higher $1/f$ and $1/f^3$ corner frequencies compared to the 70\,GHz 3GPP baseline, concentrating more PN energy within the occupied bandwidth. This exacerbates inter-carrier interference and Doppler leakage—especially under long integration times—resulting in moderate velocity degradation and significant sidelobe saturation.  

Table~\ref{tab:summary} consolidates the sensing performance across frequencies and PN conditions. While 130\,GHz delivers finer Doppler resolution in the PN-free case, the benefit diminishes under PN, particularly in sidelobe metrics, where both frequencies exhibit nearly identical saturation levels.  

Overall, FFT-based radar detection with $\mu = 5$ (480\,kHz) approaches the Cramér–Rao bound under PN-free conditions but exhibits clear error floors when realistic impairments are included. Range RMSE stabilizes around 0.04-0.05\,m, showing that accuracy is largely bandwidth-limited and only marginally affected by PN. In contrast, velocity RMSE is more sensitive: PN-free operation achieves floors of $\approx 0.15$\,m/s at 70\,GHz and $\approx 0.08$\,m/s at 130\,GHz, while PN-impaired cases flatten at $\approx 0.18$\,m/s and $\approx 0.12$\,m/s, respectively. Sidelobe suppression is most severely impacted, with PSLR saturating near $-6$\,dB and ISLR near $-4$\,dB under PN, regardless of carrier frequency.  

These results emphasize that although higher frequencies enable finer Doppler resolution, they are also more vulnerable to PN-induced distortion. This highlights the need for PN-aware waveform and numerology design in sub-THz ISAC. The proposed evaluation framework—combining hardware-calibrated PN models, standardized numerologies, and low-complexity FFT-based processing—can be readily extended to higher sub-THz bands. Such insights are critical for future multi-band ISAC transceivers, where designs must jointly account for frequency-dependent impairments, resolution requirements, and hardware constraints.

\begin{table}[h]
\centering
\caption{Summary of Sensing Performance Differences (Fixed $\Delta f = 480$ kHz)}
\label{tab:summary}
\begin{tabular}{|l|c|c|}
\hline
\textbf{Metric} & \textbf{70 GHz} & \textbf{130 GHz} \\
\hline
Range RMSE (PN-Free) & $\approx 0.04$ m & $\approx 0.04$ m \\
Range RMSE (with PN) & $\approx 0.05$ m & $\approx 0.05$ m \\
Velocity RMSE (PN-Free) & $\approx 0.15$ m/s & $\approx 0.08$ m/s \\
Velocity RMSE (with PN) & $\approx 0.18$ m/s & $\approx 0.12$ m/s \\
PSLR (PN-Free @ 30 dB) & $<-12$ dB & $<-12$ dB \\
PSLR (with PN) & $\approx -6$ dB & $\approx -6$ dB \\
ISLR (PN-Free @ 30 dB) & $\approx -7$ dB & $\approx -8$ dB \\
ISLR (with PN) & $\approx -4$ dB & $\approx -4$ dB \\
\hline
\end{tabular}
\end{table}

Overall, FFT-based radar detection at $\mu = 5$ approaches the Cramér–Rao bound in ideal conditions. PN introduces minimal impact on range, moderate degradation in velocity, and substantial sidelobe distortion. While 130 GHz improves Doppler resolution, it is more susceptible to PN-induced errors, emphasizing the importance of waveform design aligned with oscillator impairments.

The evaluation framework presented here—combining empirical PN models, fixed numerology, and low-complexity FFT-based processing—can be readily extended to other bands. These insights are particularly relevant for future multi-band ISAC systems where transceivers must adapt to varying frequency-dependent impairments, resolution needs, and hardware constraints.

While this study provides a hardware-calibrated evaluation of OFDM-based ISAC under D-band phase noise, several limitations remain. First, the analysis is restricted to 70\,GHz and 130\,GHz; extension to higher sub-THz bands (e.g., 300\,GHz) is required to fully generalize the framework. Second, the focus is on sensing performance metrics (RMSE, PSLR, ISLR), whereas communication-related metrics such as EVM, BER, and achievable rate are not considered and remain important directions for future work. Third, phase-noise mitigation techniques, including PTRS insertion, advanced CPE tracking, and AI-assisted compensation, are not evaluated here but will be essential to assess practical system robustness. Finally, the results are based on FFT-based radar processing, which is limited by grid resolution; alternative high-resolution algorithms such as MUSIC could provide further improvements.  

Addressing these limitations will allow for a more complete understanding of PN effects across sensing and communication, and guide the development of multi-band sub-THz ISAC transceivers that balance waveform design, hardware impairments, and system-level performance.

\section{Conclusion}
This paper presented a PN-aware evaluation of OFDM-based ISAC in the D-band using a hardware-tuned 3GPP model at 130\,GHz and FFT-based radar processing. With $\mu=5$ (480\,kHz), simulations showed range RMSE floors of 0.04-0.05\,m and velocity RMSE floors of 0.12-0.18\,m/s. Doppler sidelobe suppression also saturated, with PSLR $\approx -6$\,dB and ISLR $\approx -4$\,dB.  

These results confirm that range accuracy is mainly bandwidth-limited, while Doppler estimation and sidelobe behavior are strongly PN-sensitive. The findings emphasize the need for PN-aware waveform and numerology design in sub-THz ISAC. Future work will extend the framework to higher



\bibliographystyle{IEEEtran}
\bibliography{IEEEabrv,bibliography}

\begin{thebibliography}{10}
\providecommand{\url}[1]{#1}
\csname url@samestyle\endcsname
\providecommand{\newblock}{\relax}
\providecommand{\bibinfo}[2]{#2}
\providecommand{\BIBentrySTDinterwordspacing}{\spaceskip=0pt\relax}
\providecommand{\BIBentryALTinterwordstretchfactor}{4}
\providecommand{\BIBentryALTinterwordspacing}{\spaceskip=\fontdimen2\font plus
\BIBentryALTinterwordstretchfactor\fontdimen3\font minus \fontdimen4\font\relax}
\providecommand{\BIBforeignlanguage}[2]{{%
\expandafter\ifx\csname l@#1\endcsname\relax
\typeout{** WARNING: IEEEtran.bst: No hyphenation pattern has been}%
\typeout{** loaded for the language `#1'. Using the pattern for}%
\typeout{** the default language instead.}%
\else
\language=\csname l@#1\endcsname
\fi
#2}}
\providecommand{\BIBdecl}{\relax}
\BIBdecl

\bibitem{wild2021ieee}
T.~Wild, V.~Braun, and G.~Fettweis, ``Joint radar and communication in future wireless networks: A roadmap,'' \emph{{IEEE} Commun. Mag.}, vol.~59, no.~2, pp. 62--68, 2021.

\bibitem{saad2020overview}
W.~e.~a. Saad, ``A vision of {6G} wireless systems: Applications, trends, technologies, and open research problems,'' \emph{{IEEE} Signal Process. Mag.}, 2020.

\bibitem{towards_6G_AVVR}
M.~Giordani, M.~Polese, M.~Mezzavilla, S.~Rangan, and M.~Zorzi, ``Toward {6G} networks: Use cases and technologies,'' \emph{{IEEE} Commun. Mag.}, vol.~58, no.~3, pp. 55--61, 2020.

\bibitem{above100_isac}
T.~Rappaport \emph{et~al.}, ``Wireless communications and applications above 100 {GHz}: Opportunities and challenges for{ 6G} and beyond,'' \emph{IEEE Access}, vol.~7, pp. 78\,729--78\,757, 2019.

\bibitem{fundamental_RCS}
E.~M. Taghavi, H.~Saarnisaari, and M.~Juntti, ``Fundamental and practical performance assessment in monostatic {ISAC}: From {Sub-6GHz} to {Sub-THz},'' in \emph{2024 3rd International Conference on 6G Networking (6GNet)}, 2024, pp. 221--226.

\bibitem{mmwaveOFDM_radar}
M.~Mirabella, P.~D. Viesti, A.~Davoli, and G.~M. Vitetta, ``Deterministic signal processing techniques for ofdm-based radar sensing: An overview,'' \emph{{IEEE} Access}, vol.~11, pp. 68\,872--68\,889, 2023.

\bibitem{mimo_ofdm_ISAC}
Z.~Wei, J.~Piao, X.~Yuan, H.~Wu, J.~A. Zhang, Z.~Feng, L.~Wang, and P.~Zhang, ``Waveform design for mimo-ofdm integrated sensing and communication system: An information theoretical approach,'' \emph{{IEEE} Trans. Commun.}, vol.~72, no.~1, pp. 496--509, 2024.

\bibitem{understanding_effectsPN}
A.~Garcia~Armada, ``Understanding the effects of phase noise in orthogonal frequency division multiplexing ({OFDM}),'' \emph{{IEEE} Trans. Broadcast.}, vol.~47, no.~2, pp. 153--159, 2001.

\bibitem{PN_THEORICAL_model}
S.~Bicais and J.-B. Dore, ``Phase noise model selection for sub-thz communications,'' in \emph{2019 IEEE Global Communications Conference (GLOBECOM)}, 2019, pp. 1--6.

\bibitem{PN_modeling_real}
H.~Han, J.~Park, J.~Kim, K.~Bae, and I.~Na, ``Baseband phase noise modeling and analysis for 140ghz thz dft-s-ofdm system,'' in \emph{2022 IEEE Globecom Workshops (GC Wkshps)}, 2022, pp. 1748--1753.

\bibitem{8821684}
J.~Huang, Y.~Zhang, and S.~Luo, ``Analysis of pilot-aided channel estimation in {OFDM} system under phase noise,'' in \emph{Proc. IEEE ICCCS}, 2019, pp. 483--487.

\bibitem{8690852}
Y.~Qi, M.~Hunukumbure, H.~Nam, H.~Yoo, and S.~Amuru, ``On the phase tracking reference signal ({PT-RS}) design for {5G} new radio (nr),'' in \emph{Proc. IEEE VTC-Fall}, 2018, pp. 1--5.

\bibitem{subcarriier_modulation_specification}
T.~Levanen, O.~Tervo, K.~Pajukoski, M.~Renfors, and M.~Valkama, ``Mobile communications beyond 52.6 {GHz}: Waveforms, numerology, and phase noise challenge,'' \emph{{IEEE} Wireless Commun.}, vol.~28, no.~1, pp. 128--135, 2021.

\bibitem{study_numerol_V2X}
S.~Khabaz, K.~O. Boulila, T.~M. Trang~Nguyen, G.~Pujolle, M.~El~Aoun, and P.~B. Velloso, ``A comprehensive study of the impact of 5g numerologies on v2x communications,'' in \emph{2022 13th International Conference on Network of the Future (NoF)}, 2022, pp. 1--9.

\bibitem{Tunned_PN}
C.~T. Parisi, S.~Badran, P.~Sen, V.~Petrov, and J.~M. Jornet, ``Modulations for terahertz band communications: Joint analysis of phase noise impact and {PAPR} effects,'' \emph{IEEE Open Journal of the Communications Society}, vol.~5, pp. 412--429, 2024.

\bibitem{keskin2022ofdm}
M.~U. Keskin, F.~Liu, and A.~L. Swindlehurst, ``Monostatic sensing with {OFDM} under phase noise: Characterization and mitigation,'' \emph{arXiv preprint}, 2022.

\bibitem{sen2023sensors}
S.~Sen, H.~Pan, Z.~Popovic, M.~A. Beach, T.~A. Tsiftsis, and I.~B. Djordjevic, ``Joint communication and radar sensing: {RF} hardware opportunities and challenges,'' \emph{Sensors}, vol.~23, no.~18, p. 7673, 2023.

\bibitem{wang2021dfts}
Y.~Wu, F.~Lemic, C.~Han, and Z.~Chen, ``Sensing integrated {DFT-Spread OFDM} waveform and deep learning-powered receiver design for terahertz integrated sensing and communication systems,'' \emph{{IEEE} Trans. Commun.}, vol.~71, no.~1, pp. 595--610, 2023.

\end{thebibliography}

\end{document}